\documentstyle[12pt]{article}
\title{Roper resonance in the relativistic three-quark model}
\author{D.V.Ivanov }
\date{}
\begin{document}
\maketitle
\centerline{Department of Theoretical Physics, St. Petersburg }
\centerline{State University 198904, St.Petersburg, Russia}
\vspace{2cm}

\begin{abstract}
The relativistic three-quark equations are found in the framework of the
dispersion relation technique. The approximate solutions of these equations
using the method based on the extraction of leading singularities of the
amplitude are obtained. The calculated mass values of two ($ N=2, 56^* $)
baryonic multiplets $ J^P ={\frac{1}{2}}^{+},{\frac{3}{2}}^{+} $ are in good
agreement with the experimental ones.
\end{abstract}

In the recent papers [1,2] the Faddeev equations are represented in the form
of the dispersion relation over the two-body subenergy. The behaviour of the
low-energy three-body amplitude is determined by its leading singularities
in the pair invariant masses. Then the purpose was to extract the singular
part of the amplitude. The suggested method of approximate solution of the
Faddeev equations was verified on the example of the S-wave baryonic
spectroscopy. We calculated the lowest baryon masses ($ J^P ={\frac{1}{2}}^
{+},{\frac{3}{2}}^{+} $) using the method based on the extraction of leading
singularities of the amplitude.

In the present paper the relativistic Faddeev equations are also constructed
in the form of the dispersion relation over the two-body subenergy. We
calculated the mass values of two ($N=2, 56^* $) baryon multiplets with
$ J^P ={\frac{1}{2}}^{+},{\frac{3}{2}}^{+} $, which are in good agreement
with the experimental ones [3].

One used the results of the bootstrap quark model [4] and determined the
diquark amplitude with $ J^P =0^{+}, 1^{+}$. The integral equation systems,
corresponding to the ($N=2, 56^* $) baryonic multiplets $ J^P
={\frac{1}{2}}^{+},{\frac{3}{2}}^{+} $ are analogous to the integral
equations for the S-wave lowest baryonic multiplets $ J^P ={\frac{1}{2}}^{+},
{\frac{3}{2}}^{+}$ [1,2]. However, for the excited baryons the long-range
forces, which  due to the confinement, are important. Namely, the
box-diagrams can be important in the formation of hadron spectra [5]. For the
sake of simplicity we restrict ourselves to the introduction of the quark
mass shift $\Delta $, which are defined by the contributions of the nearest
production thresholds of pair mesons and baryons. We suggested that the
parameter $\Delta $ takes into account the confinement potential effectively:
$m_{eff}=m+\Delta $ and $m^{s}_{eff}=m^{s}+\Delta $ and changes the behaviour
of diquark amplitude [6]. It allows to construct the excited baryonic
amplitudes and calculate the mass spectrum by analogy with [6]. In the
considered calculation the quark masses ($m$ and $m^s$) are analogous [1,2]:
$m$=0.410~GeV, $m^s$=0.557~GeV. $\Delta $ is equal 0.168~GeV.

The construction of the approximate solutions of Faddeev equations is based
on the extraction of the leading singularities which are close to region of
pair energy $\sim 4m^2 $. First of all there are threshold square root
singularities. Also possible are pole singularities which correspond to the
bound states. They are situated on the first sheet of complex pair energy
plane in the case of real bound state and on the second sheet in case of
virtual bound state. This diagrams have only two-particle singularities.
Other diagrams apart from two-particle singularities have their own specific
triangle singularities. It is weaker than two-particle singularities. Such
classification allows us to search the approximate solution by taking into
account some definite number of leading sinfularities and neglecting all the
weaker ones. We consider the approximation, which corresponds to the single
interaction of all three particles (two-particle and triangle singularities)
[1,2].

In the present paper the suggested method of approximate solution of the
relativistic three-quark equations allows us to calculate the excited
baryons spectrum ($N=2, 56^* $) with $ J^P ={\frac{1}{2}}^{+},{\frac{3}
{2}}^ {+}$. The interactions, determined this spectrum, are similar to ones in
the S-wave lowest baryons case. We use two vertex constants $g_0 =0.702$ and
$g_1 =0.540$, which corresponds to the quark-quark interaction in $0^{+}$
and $1^{+}$ states. One introduce two dimensionless cut-off parameters
${\lambda}_0^s =10.5$ and ${\lambda}_1 =11.5$, which can be determined by
mean of fixing of excited baryon mass values ($N^{*}, {\Sigma}^{*}$). We
decreased the colour-magnetic interaction in $0^{+}$ strange channels
(${\lambda}_0^s < {\lambda}_1$). Then we calculated the mass values of two
excited multiplets $ J^P ={\frac{1}{2}}^{+},{\frac{3}{2}}^{+}$ ($N=2,
56^* $) (Table) in good agreement with the experimental data [3] and
the other model results [7-12].

\begin{center}

Table

Masses of excited ($N=2, 56^* $) baryon resonances multiplets
$ J^{P}={\frac{1}{2}}^{+},{\frac{3}{2}}^{+}$.
\vspace{0.5cm}

\begin{tabular}{|l|l|l|l|}             \hline
$J^{P}={\frac{1}{2}}^{+}$ & M(GeV)    & $J^{P}={\frac{3}{2}}^{+}$  & M(GeV) \\
\hline
$N^{*}$          & 1.440(1.440) & ${\Delta}^{*}$           & 1.715(1.600)
\\ \hline
${\Lambda}^{*}$  & 1.610(1.600) & ${\Sigma}^{*}$           & 1.865
\\ \hline
${\Sigma}^{*}$   & 1.655(1.660) & ${\Xi}^{*}$              & 2.010
\\ \hline
${\Xi}^{*}$      & 1.785        & ${\Omega}^{*}$           & 2.155   \\ \hline
\end{tabular}
\end{center}

\noindent
The cut-off parameters ${\lambda}_0^s =10.5 $ and ${\lambda}_1 =11.5 $. The
vertex functions $g_0 =0.702, g_1 =0.540$. Experimental values of the baryon
masses [3] are given in parentheses.

\vspace{0.5cm}
The reason of essential difference between ${\Sigma}^{*}$ and ${\Lambda}^{*}$
is the spin of the lighter diquark. The model explain both the sign and 
magnitude of this mass splitting.

The Roper resonance $ N^{*}(1440) $ is the lightest member of the $ J^P =
{\frac{1}{2}}^{+} $ multiplet. Within the constituent quark picture [12,13]
this resonance commonly assigned to a radial excitation of the nucleon,
whereas it has been argued [14-16] that it might be a hybrid state,
containing an explicit excited glue-field configuration (i.e. gqqq-state).

The model under consideration proceeds from the assumption that the quark
interaction forces are the two-component ones. The long-range component of
the forces is neglected. The creation of low-lying baryons is mainly due to
the constituent gluon exchange. But for the excited baryons the long-range
forces are important. The confinement with comparatively large energy is
actually realized as the production of the new $ q\bar q $ pairs. The
long-range forces are determined by the contribution of the nearest
production thresholds of pair quarks. We suggest that quark mass shift
$\Delta $ takes into account the confinement potential effectively and
changes the behaviour of pair quarks amplitude. It allows us to construct
the excited baryon amplitudes and calculate the baryon mass spectrum by
analogy with P-wave meson spectrum in the bootstrap quark model [6].

We manage with the quarks as with real particles. However, in the soft
region, the quark diagrams should be treated as spectral integrals over
quark masses with the spectral density $\rho (m^2)$: the integration over
quark masses in the amplitude puts away the quark singularities and introduces
the hadron ones. We can believe that the approximation $\rho (m^2) \to
\delta (m^2 -m_{eff}^{2})$ could be possible for the excited baryons (here
$m_{eff}$ is the effective ''mass'' of the constituent quark). We hope this
approach is sufficiently good for the calculation of excited baryonic
spectrum.

The author would like to thank S.M.~Gerasyuta for the main idea of this
investigation and useful discussions.


\begin{thebibliography}{16}

\bibitem{1}
S.M. Gerasyuta, Nuovo Cim. A106 (1993) P.37
\bibitem{2}
S.M. Gerasyuta, Z.Phys. C60 (1993) P.683
\bibitem{3}
Particle Data Group, Phys.Rev. D54 (1996) P.1
\bibitem{4}
V.V. Anisovich, S.M. Gerasyuta, A.V. Sarantsev, Int.J.Mod.Phys. A6 (1991) P.625
\bibitem{5}
V.V. Anisovich, Proc.Int.Simp. Pion-Nucleon, Nucleon-Nucleon Physics. Gatchina.
V.2 (1989) P.237
\bibitem{6}
S.M. Gerasyuta, I.V. Keltuyala, Sov.J.Part.Nucl. V.54 (1991) P.793
\bibitem{7}
A.J.G. Hey, R.L. Kelly, Phys.Rep. V.96 (1983) P.72
\bibitem{8}
M. Jones et al., Nucl. Phys. B129 (1977) P.45
\bibitem{9}
I.M. Barbour, D.K. Ponting, Z.Phys. C4 (1980) P.119
\bibitem{10}
N. Isgur, G. Karl, Phys.Rev. D19 (1979) P.2653
\bibitem{11}
N. Isgur, R. Koniuk, Phys.Rev.Lett. V.44 (1980) P.845
\bibitem{12}
N. Isgur, R. Koniuk, Phys.Rev. D21 (1980) P.1868
\bibitem{13}
S. Capstick, B. Keister, Phys.Rev. D51 (1995) P.3598
\bibitem{14}
T. Barnes, F.E. Close , Phys.Lett. B123 (1983) P.89
\bibitem{15}
Z. Li, F.E. Close, Phys.Rev. D42 (1990) P.2207
\bibitem{16}
F. Cardarelli, E. Pace, G. Salm\`{e}, S. Simula, Phys.Lett. B397 (1997) P.13
\end{thebibliography}
\end{document}